\documentclass[aps,notitlepage]{revtex4-2} 

\usepackage{graphicx} 
\usepackage{subcaption}  
\usepackage{amsmath}
\usepackage{amssymb}
\usepackage{latexsym}
\usepackage{xcolor,soul}

\newcommand{\be}{\begin{equation}}
\newcommand{\ee}{\end{equation}}
\newcommand{\bea}{\begin{eqnarray}}
\newcommand{\eea}{\end{eqnarray}}

\begin{document}

\title{Peccei-Quinn Genesis}

\author{Eung Jin Chun}
\affiliation{Korea Institute for Advanced Study, Seoul 130-722, Korea}

\author{Hyun Min Lee and  Jun-Ho Song}
\affiliation{Department of Physics, Chung-Ang University, Seoul 06974, Korea}

\begin{abstract}
We propose a cogenesis mechanism that unifies the origin of QCD axion dark matter and the baryon asymmetry of the Universe in the framework of Peccei-Quinn pole inflation. The model integrates the Peccei-Quinn symmetry with the seesaw mechanism for neutrino masses. This allows for spontaneous leptogenesis, which generates the required $B-L$ asymmetry around the seesaw scale. The necessary initial axion kinetic misalignment is naturally sourced by a PQ field driving pole inflation. Analysis within the KSVZ axion model demonstrates that achieving simultaneous correct DM abundance and baryon asymmetry limits the axion decay constant to be smaller than about $10^9$ GeV. This framework offers a unified solution to four fundamental problems: the strong CP problem, neutrino mass, matter-antimatter asymmetry, and inflation.

\end{abstract}

\maketitle

\section{Introduction}

The QCD axion is a Goldstone boson of a spontaneously broken Peccei-Quinn (PQ) symmetry $U(1)_{PQ}$ that has a color anomaly. It has been introduced to solve the strong CP problem \cite{pq77,ww78,ksvz79,dfsz81} as its vacuum expectation value, and thus the effective $\theta$ term settles down to zero due to the QCD potential. As a consequence, the axion obtains a tiny mass $m_a \propto 1/f_a$ where $f_a$ is the axion decay constant restricted to the range $f_a \sim 10^{8-12}$ GeV by various observations. For the KSVZ axion \cite{ksvz79}, which couples  almost only to protons, the strongest lower bound can be obtained from SN1987A: $f_a>4\times 10^8$ GeV \cite{cr24}. The upper limit is set by the axion dark matter abundance, which is produced by the standard misalignment mechanism \cite{misal83}. On the other hand, one can rely on the kinetic misalignment \cite{chh19} as the origin of the dark matter density even for a lower value of $f_a$. 

The axion kinetic misalignment can also be the source of the observed baryon asymmetry through the mechanism of spontaneous baryogenesis \cite{ck87}, as it can provide a CPT violating background while the electroweak sphaleron interaction, that violates $B+L$ symmetry,  is in equilibrium \cite{ch19}.  The final baryon asymmetry is determined at the electroweak phase transition below which the sphaleron interaction decouples. However, it conflicts with the dark matter abundance from  standard QCD axions.  

To resolve this difficulty, we consider a simple extension of the PQ symmetry to include the seesaw mechanism that explains the observed neutrino masses and mixing \cite{seesaw77}, which involves lepton number violation. This enables spontaneous leptogeneis that generates $B-L$ asymmetry around the seesaw scale \cite{cj23,cls25}.  Adjusting this scale, we can realize the cogenesis of baryon asymmetry and dark matter of the Universe.

Another key ingredient of spontaneous leptogenesis is the origin of the initial kinetic misalignment, which could be related to inflation dynamics \cite{ad85}.  In this regard, it appears natural to explore a minimal setup where the PQ field, breaking the PQ symmetry, drives inflation. Its radial degree is supposed to play a role of the inflaton and the axial degree (axion) develops the initial kinetic motion at the end of inflation through a higher-dimensional operator violating the PQ symmetry explicitly \cite{hmlee}.  We will examine the cogenesis condition in the KSVZ model and analyze the relevant inflation dynamics in the framework of pole inflation. Then, we find out the allowed parameter regions which turns out to be quite limited.  On the other hand, the DFSZ framework \cite{dfsz81} is not considered, since  reheating through PQ field decay into Higgs doublets is too inefficient to achieve cogenesis.

\section{Cogenesis requirements}

\subsection{Spontaneous leptogenesis and dark matter abundance}

Consider a field $\Phi$ that carries the PQ charge $+1$ and breaks $U(1)_{PQ}$ spontaneously. After breaking the symmetry, it is represented by $\Phi=\frac{v_a}{\sqrt{2}}  e^{i {\theta}}$ where $\theta\equiv a/v_a$. Given the kinetic misalignment $\dot\theta$ at a temperature $T$,  the PQ number normalized by the entropy density is a conserved quantity:
\begin{equation}
    Y_\theta \equiv {n_\theta \over s}=\frac{v_a^2 \dot\theta}{{2\pi^2\over 45} g_* T^3}
\end{equation}
In the background of $\dot\theta$, neutrino Yukawa interactions, $y_\nu l N H$ with an RHN $N$, may be in equilibrium to generate a non-vanishing $B-L$ asymmetry: $Y_{B-L} \equiv {n_{B-L}\over s}$  where $n_{B-L} = {1\over6} \mu_{B-L} T^2$ and the chemical potential $\mu_{B-L} \propto \dot\theta$ is determined by the equilibrium conditions of a given system.  Thus, one can parameterize the resulting baryon asymmetry $Y_B$ by 
\begin{equation} \label{PQtoB}
    Y_B=c_B Y_\theta \left( T_{B-L}\over v_a \right)^2
\end{equation}
where $c_B \equiv c_{\rm sph} {\mu_{B-L} \over 6 \dot\theta}$ is a model-dependent constant, and $T_{B-L}$ is the temperature at which the final $B-L$ asymmetry is determined. Here, $c_{\rm sph}$ is the usual $(B-L)$-to-$B$ conversion factor which is given by $c_{\rm sph}={28\over 79}$ when all the Yukawa interactions are in equilibrium (namely, there is no flavor dependence). 

The neutrino Yukawa interactions violating lepton number in our case are the decay and inverse decay of $N$, and   $T_{B-L}$ is the temperature at which the inverse decay goes out of equilibrium. This occurs in the so-called strong washout regime, where the decay rate $\Gamma_D$ of $N$ is higher than the Hubble rate at temperature $T=M_N$: $K\equiv \Gamma_D/H(M_N) > 1$. For our analysis, we stick to the case of full equilibration with $K=50$ \cite{cls25}.

Requiring $Y_B \approx 0.87 \times 10^{-10}$, we determine $Y_\theta$ by (\ref{PQtoB}). If $Y_\theta$ is also the main source of dark matter density, the following condition must hold \cite{ch19}: 
\begin{equation}
   2 m_a |Y_\theta| \approx 0.44 \mbox{ eV} .\label{dark matter abundance}
\end{equation}
Following the convention of the anomaly coupling of the axion $a$ to the gluon fields:
\begin{equation}
    -{\cal L}_{aGG}= {1\over 32 \pi^2} {a \over f_a} G \tilde G ,
\end{equation}
we get the relation $v_a = N_{\rm DW} f_a$ where $N_{\rm DW}$ is the QCD anomaly coefficient which we will also denote as $c_S$.
Then, the axion mass is determined to be \cite{bor16}
\begin{equation} \label{mafa}
    m_a \approx 5.7 \mbox{meV} \left( 10^{9} \mbox{GeV} \over f_a\right).
\end{equation}
Thus, a successful cogenesis requires that
\begin{equation} \label{Eqcogen}
    2.25 {N_{DW}^2\over |c_B|} \left(f_a \over 10^9 \mbox{GeV}  \right) \left( 10^3 \mbox{GeV} \over M_N /z_N \right)^2 =1 .
\end{equation}
In the following subsections, we will examine the KSVZ axion model that incorporates the seesaw mechanism for neutrino masses and determine the coefficient $c_B$ appearing in the baryon asymmetry in (\ref{PQtoB}).

\subsection{KSVZ axion and seesaw mechanism}

The KSVZ axion is implemented by the existence of extra quarks $(Q, Q^c)$. In our scheme, RHNs couple to the PQ field $\Phi$ and thus we have  
\begin{equation}
    -{\cal L}_{\rm KSVZ} = y_Q \Phi Q Q^c + \frac{1}{2} y_N \Phi N N + h.c. \label{KSVZ yukawa}
\end{equation}
where $\Phi=\frac{v_a}{\sqrt{2}}  e^{ia/v_a}$ is the PQ field (with the PQ charge $+1$) after the PQ symmetry breaking.  and $Q Q^c$ is a Dirac fermion bilinear with the mass $M_Q= \frac{y_Q}{\sqrt{2}}  v_a$ and $N$ gets the Majorana mass $M_N= \frac{y_N}{\sqrt{2}}v_a$ after rotating away the phase field $e^{i a/v_a}$. Note that we use the chiral notation for the fermions.  At the energy scale below $M_Q$, the Standard Model (SM) Lagrangian includes
\begin{equation} \label{SMeq}
\begin{array}{lcl} 
    -{\cal L}_{\rm Yukawa}  &=& y_u q u^c H +y_d q d^c \tilde{H} + y_e l e^c \tilde{H} +y_\nu l N H + h.c.  \\
   -{\cal L}_{\rm anomaly} &=&  {c_S \over 32 \pi^2} {a\over v_a} G\tilde{G} + {c_W \over 32 \pi^2} {a\over v_a} W\tilde{W} 
\end{array}
\end{equation}
where the chiral notation is also used for the quark (lepton) doublets $q\, (l)$ and singlets $u^c,d^c (e^c)$, and the Higgs doublet and its conjugate are denoted by $H$ and $\tilde{H}$. Here we suppressed the family indices which will be recovered whenever necessary.  In the anomaly term, we keep only the axion couplings to the $SU(3)_c$ and $SU(2)_L$ gauge bosons that have the anomaly coefficients $c_S$ and $c_W$, respectively.  In the KSVZ model, $c_S=N_{DW}$ and $c_W$, are given by the total number of heavy quarks and the number of heavy quark doublets, respectively. 

Assuming all the Yukawa and sphaleron interactions are in equilibrium, one can generalize the calculation in Ref.~\cite{DEMY20} to include the seesaw sector. 
As  the first generation quark Yukawa interactions are much weaker than the others as well as the strong and weak sphaleron interactions: $\Gamma_{y_{u_1, d_1}}\ll \Gamma_{y_{u_{2,3},d_{2,3}}}, \Gamma_{SS,WS}$, the equilibrium conditions are satisfied by the following chemical potential relations involving only quarks:
\begin{equation}
\begin{array}{l}
    \mu_{q_i}+ \mu_{u^c_i}+\mu_H = c_S {m_d^2 \over m_u^2 + m_d^2} \dot\theta \,\delta_{i1}, \\
     \mu_{q_i}+ \mu_{d^c_i}-\mu_H = c_S  {m_u^2 \over m_u^2 + m_d^2}\dot\theta \,\delta_{i1}, \\
     \sum_{i} \left( 2\mu_{q_i}+\mu_{u^c_i}+ \mu_{d^c_i} \right)=c_S \dot\theta.
\end{array}
\end{equation}
On the other hand, the equilibrium condition involving leptons requires 
\begin{equation}
    \begin{array}{l}
         \mu_{l_i}+\mu_{e^c_i}-\mu_H=0, \\
         \mu_{l_i}+\mu_H=x_l \dot\theta, \\
         \sum_i \left( 3 \mu_{q_i} + \mu_{l_i}\right) = c_W \dot\theta,
    \end{array}
\end{equation}
with no flavor dependence: $\mu_{l_i}=\mu_l$, etc. Note that the second equation comes from the shift in the chemical potential $\mu_l \to \mu_l - x_l \dot\theta$ as leptons carry PQ charges, and thus their charge operators couple to $\dot\theta$.

Therefore, we find the chemical potential for the $B-L$ charge ($q_{B-L}={1\over6} \mu_{B-L} T^2$): 
\begin{equation} \label{BLksvz}
    \mu_{B-L} = {1\over 33} \left( 28 c_W - c_S {57 m_u^2 -15 m_d^2 \over m_u^2 + m_d^2 } -153 x_l \right) \dot\theta
\end{equation}
where $x_l$ is the PQ charge of the lepton doublets $l_i$ ($x_l={1\over2}$  in our case).
This results in 
\begin{equation}
    c_B = {28 \over 79} {1\over 198}\left(28 c_W - c_S {57 m_u^2 -15 m_d^2 \over m_u^2 + m_d^2 } -153 x_l\right).
\end{equation}
Taking $c_S=N_{DW}=1$, $c_W=0$, $x_l=1/2$, and  $m_u/m_d=0.462$ \cite{pdg}, we find $c_B=-0.13$.

Let us finally comment on the coupling of the PQ field to the Higgs doublet: $\lambda_{\Phi H} |\Phi|^2 |H|^2$. To avoid  fine-tuning in the Higgs sector, one may require $\lambda_{\Phi H} < \langle H\rangle^2/\langle \Phi\rangle^2$, which is smaller than about $10^{-14}$. This coupling is too small to provide a sufficient reheat in PQ inflation scenarios, which will be discussed in the following section.  

Given $c_B=-0.13$, we can determine the value of $M_N$ or the Yukawa coupling $y_N=\sqrt{2}M_N/v_a$ for cogenesis (\ref{Eqcogen}):
\begin{equation}
    y_N = 5.9 \times 10^{-6} z_N \sqrt{ {10^9 {\rm GeV} \over f_a}}. \label{yN}
\end{equation}
From the analysis in \cite{cls25}, we find $z_N\approx 10$  for $K=50$, which is assumed in our numerical results.

\section{PQ pole inflation and axion kinetic misalignment}

Let us begin by recapitulating the main features of the PQ inflation at the pole that we use for our purpose.
In pole inflation, the inflaton's kinetic term becomes singular at a specific point, which effectively stretches the field space. When the inflaton field approaches the pole, the redefinition of the field to make its kinetic term canonical results in an effectively flat potential \cite{pole,pole2}. Thus, a wide range of different underlying potentials, when subjected to the pole in the kinetic term, can drive inflation and lead to very similar predictions for the observable cosmological parameters such as the spectral index $n_s$ and the tensor-to-scalar ratio $r$.  

In the PQ mechanism, $U(1)_{\rm PQ}$ is broken by a complex (PQ) field $\Phi$ typically by a quartic potential
\bea \label{VPQC}
V_{\rm PQC}=\lambda_\Phi \bigg(|\Phi|^2 -\frac{v_a^2}{2}\bigg)^2,
\eea
 which we take as an underlying inflaton potential where the radial mode of $\Phi$ acts as the inflaton. 
The axial mode of $\Phi$ is the axion whose kinetic misalignment is supposed to generate the baryon asymmetry of the universe and then turns into dark matter.  For this purpose, we introduce a higher-dimensional operator that softly violates the PQ symmetry \cite{hmlee}:
\begin{equation} \label{VPQV}
    V_{\rm PQV}= \lambda_n\frac{\Phi^n}{M_P^{n-4}}+\text{h.c.}.
\end{equation}
 which is supposed not to perturb the features of inflation driven by (\ref{VPQC}),  and the axion solution to the strong CP problem \cite{axionquality}.

\subsection{PQ pole inflation}

We consider the Jordan frame Lagrangian for the PQ field $\Phi$ with a non-conformal coupling
\bea
\frac{{\cal L}_J}{\sqrt{-g_J}}= -\frac{M_P^2}{2} \Omega(\Phi) R+ |\partial_\mu\Phi|^2 - \Omega^2 V_E.
\eea
where $V_E=V_{\rm PQC}+V_{\rm PQV}$ and the non-minimal coupling function is taken to be
\bea
\Omega(\Phi) &=& 1-2 \frac{|\Phi|^2}{\Lambda^2}.
\eea
Here, the new cutoff-scale $\Lambda$ in the non-minimal coupling is related to a parameter $\zeta\equiv M_P^2/\Lambda^2$, which was taken to a  conformal coupling, $\zeta=\frac{1}{6}$, in Ref.~\cite{hmlee}. 

We can go to the Einstein frame by making a Weyl transformation: 
$g_{J.\mu\nu}=g_{E,\mu\nu}/\Omega$:
\bea 
\frac{{\cal L}_E}{\sqrt{-g_E}}&=&-\frac{M_P^2}{2}R+\frac{1}{\Omega}\, |\partial_\mu\Phi|^2 +\frac{3}{4}M_P^2 \frac{(\partial_\mu\Omega)^2}{\Omega^2}-V_E   \nonumber \\
&=&-\frac{M_P^2}{2}R +\frac{1+\zeta(6\zeta-1)\frac{\rho^2}{M_P^2}}{2(1-\zeta \frac{\rho^2}{M_P^2})^2}\,(\partial_\mu\rho)^2 +\frac{\rho^2(\partial_\mu\theta)^2}{2(1-\zeta\frac{\rho^2}{M_P^2})} - V_E  \label{Lagrangian E}
\eea
where we used the notation $\Phi=\frac{1}{\sqrt{2}}\rho\,e^{i\theta}$ in the second line. The potential $V_{E}= V_{\rm PQC}+V_{\rm PQV}$ is explicitly given by
\begin{align} \label{PQCV}
    V_{\rm PQC}=&\frac{\lambda_{\Phi}}{4}(\rho^2-v_a^2)^2, \nonumber \\  
    V_{\rm PQV}=& \frac{|\lambda_n|}{\sqrt{2^{n-2}}}\frac{\rho^n}{M_P^{n-4}} \cos{(n\theta+\delta)}.
\end{align}
Here, $\delta$ is the phase of a complex parameteer $\lambda_n$.

One can see that a pole appears at $\rho=M_P/\sqrt{\zeta}=\Lambda$ around which inflation is occurring. To get the canonically normalized inflaton field $\phi$, we make the field redefinition $\rho=\rho(\phi)$ satisfying 
\bea
\frac{d\rho}{d\phi}= {1\over \sqrt{\Xi(\rho)}}  \equiv  \frac{(1-\zeta\frac{\rho^2}{M_P^2})}{\sqrt{1+\zeta(6\zeta-1)\frac{\rho^2}{M_P^2}}}, \label{Xi}
\eea
where $\phi$ is the canonical normalized real scalar.
In the limit $\rho \rightarrow \frac{M_P}{\sqrt{\zeta}}$ during inflation, we can approximately get the canoncial normalized field for $\zeta \gtrsim 1$ as
\bea
\frac{d\rho}{d\phi}&\simeq& \frac{(1-\zeta\frac{\rho^2}{M_P^2})}{\sqrt{6\zeta}}, \nonumber \\
\rho &\simeq & \frac{M_P}{\sqrt{\zeta}}\tanh{\left(\frac{\phi}{\sqrt{6}M_P}\right)}.\label{tanh}
\eea
We rewrite the PQ field Lagrangian in terms of $\phi$ and $\theta$,  where  $\rho$ is understood to be a function of $\phi$, $\rho(\phi)$:
\begin{align}
 {\cal L}_{\rm {PQ}}=  {1\over2} (\partial_\mu\phi)^2 +{1\over2} {\rho^2\over \Omega} (\partial_\mu\theta)^2 -V_{E}(\rho,\theta)
\end{align}
where $\Omega =1-{\zeta} {\rho^2 \over M_P^2}$. Then, dynamics of the inflaton $\phi$ and the axion $\theta$ is governed by
\begin{equation} \label{DynaEq}
    \begin{split}
&\Ddot{\phi}+3H\Dot{\phi}
- \frac{\rho}{\Omega^2} 
{d\rho \over d\phi}\, \Dot{\theta}^2 + \frac{\partial V_E}{\partial \rho}\frac{d\rho}{d\phi}
=  0, \\
&\Ddot{\theta}+3H\Dot{\theta}+
 \frac{2}{\rho \Omega} {d\rho \over d\phi} \,
\Dot{\phi}\Dot{\theta}  
+{\Omega\over \rho^2}  \frac{\partial V_{E}}{\partial \theta} 
= 0  
    \end{split}
\end{equation}
where the Hubble parameter is determined by
$$
H^2 = \frac{1}{3M_P^2}\left(\frac{1}{2} {\dot\phi}^2 
+{1\over2} {\rho^2\over \Omega} \dot\theta^2 +V_E \right)  .
$$
During inflation around the pole, the inflaton energy density and the slow-roll parameters are assumed to be dominated by $\phi$, that is,  $V_E \simeq V_{\rm PQC} \gg V_{\rm PQV}$, $\partial V_{\rm PQC}/\partial \phi \gg \partial V_{\rm PQV}/\partial \phi$, as well as
\bea \label{lamninfl}
  \epsilon_\phi =\frac{M_P^2}{2 V_E^2}  \left(\frac{\partial V_{E}}{\partial \rho}\frac{d\rho}{d\phi}\right)^2 ~~\gg ~~
  \epsilon_\theta = \frac{M_P^2}{2 V_E^2} {\Omega \over \rho^2} \left(\frac{\partial V_{E}}{\partial \theta}\right)^2 
  \label{inflation condition} .
\eea
Then, the inflation energy $V_I$ and the slow-roll parameter $\epsilon_*\simeq \epsilon_{\phi *}$ at the horizon exit are 
\begin{equation}
    V_I \simeq {1\over 4} {\lambda_\Phi \over \zeta^2} M_P^4,\qquad \epsilon_* \simeq {3\over 4 N^2}
\end{equation}
where $N$ is the number of efolding. 
Thus, the CMB normalization tells us that 
\begin{equation}
    A_s=\frac{1}{24\pi^2\epsilon_*}\frac{V_I}{M_P^4}=2.1\times 10^{-9}  ~~\Rightarrow~~ 
    \frac{\lambda_\Phi}{\zeta^2}\simeq 4\times 10^{-10}
\end{equation}
for $N=60$.  
Inflation ends when $\epsilon \simeq \epsilon_\phi =1$,
 corresponding to $\rho=\rho_e$:
 \begin{equation}
      \rho_{e}\approx {0.732 \over \sqrt{\zeta}} M_P =0.732\Lambda, 
 \end{equation}
 in the regime  $\zeta\gg1$. 
 Applying the slow-roll condition until the end of inflation, the evolution equation (\ref{DynaEq}) simplifies to
 \begin{equation}
3H\Dot{\phi} + \frac{\partial V_E}{\partial \phi}
\approx  0 ,\quad
3H\Dot{\theta} 
+{\Omega\over \rho^2}  \frac{\partial V_{E}}{\partial \theta} 
\approx  0 , 
\end{equation}
 from which we find  $H\approx \sqrt{V_{\rm PQC}/3M_P^2}$, and the axion number density  at the end of inflation is given by
 \begin{equation}
     n_{\theta, e}=\frac{\rho^2_e}{\Omega_e}\, {\dot\theta}_e \approx -{1\over 3 H} {\partial V_{\rm PQV} \over \partial \theta} \Big|_{\rho=\rho_e} \label{n_theta e},
 \end{equation}
with $V_{\rm PQV}$ given in (\ref{PQCV}).  
   Thus, $n_\theta$ is determined by the coupling parameter $\lambda_n$ and the dimensionality $n$ of $V_{\rm PQV}$, which are supposed to satisfy the inflation conditions in and above (\ref{lamninfl}).  They are also bounded by the condition that  the quality of the axion solution to the strong CP problem should not be spoiled  \cite{axionquality}, which reads,  in terms of the mass-squared induced by $V_{\rm PQV}$ at $\rho=v_a$, 
\begin{equation}
 \delta m_a^2 \equiv  n^2 \frac{|\lambda_n|}{\sqrt{2^{n-2}}}\frac{v_a^{n-2}}{M_P^{n-4}}  \lesssim 10^{-10} {n\over N_{DW}} m_a^2,
\end{equation}
where $m_a$ is given by (\ref{mafa}).  We find that this condition is much milder than the inflation conditions in the parameter space allowing the cogenesis. For the purposes of our analysis, we normalize all sinusoidal functions of $\theta$ to unit amplitude, and consider the case of $v_a=f_a$ ($N_{DW}=1$).

\subsection{Post-inflation evolution: kinetic misalignment and reheating}

After the end of inflation, the axion number $a^3 n_\theta$ in a volume $a^3$ is approximately conserved  since the potential $V_{\rm PQV}$ becomes negligible as the inflaton field $\phi$ (or $\rho$) continues to roll down the potential $V_{\rm PQC}$.
That is, $n_\theta$ at the scale $a$ is
\begin{align}
    n_{\theta}= n_{\theta,e}\left(\frac{a_{e}}{a}\right)^3 .
\end{align} 
The inflaton also starts a damped oscillation around the minimum of the potential. During this epoch, the oscillation amplitude of $\phi$ and the Hubble parameter decrease due to the Hubble friction. The factor, $\Xi(\rho)\equiv \left(d\rho/d\phi\right)^{-2}$ in the kinetic term driving the pole inflation, converges to 1 when $\rho \ll  \rho_c  \equiv \frac{M_P}{\sqrt{6}\zeta}$ for which $\rho$ becomes a canonically normalized field, $\rho \approx \phi$. In this era, the potential $V_{\rm PQC}$ scales like radiation: $V_{\rm PQC} \propto 1/a^4$ \cite{inflatonosc}.
For the case of our interest, $\rho_c$ is comparable to $\rho_e$, and thus the universe approaches quickly to the radiation-like era when reheating occurs. To quantify the impact of this intermediate era, we solve numerically  the equation of motion 
in terms of the non-canonical field  $\rho$:
\begin{align}
    \Ddot{\rho} + 3H \dot{\rho} + \frac12 {1\over \Xi}  \frac{d \Xi}{d\rho} \Dot{\rho}^2 + \frac{1}{\Xi}\frac{dV_{\rm PQC}}{d\rho} \simeq 0,\qquad
    H^2=\frac{1}{3M_P^2}\left[ \frac{1}{2}\Xi \Dot{\rho}^2+V_{\rm PQC} \right],
     \label{e.o.m of rho}
\end{align}
driven from (\ref{Lagrangian E}, \ref{Xi}).  From this, we obtain the numerical value of the quantity $c_N$ as shown in detail in the appendix:
\begin{equation}
c_N \equiv { V_{\rm PQC}(\rho_c) a_c^4 \over V_{\rm PQC}(\rho_e) a_e^4} \simeq 0.37 \zeta +1.79\,, \label{cN}
\end{equation}
which allows us to parametrize the axion number density at the scale factor $a>a_c$, (or $\rho=\phi <\rho_c$), as 
\begin{equation}
    n_\theta (\phi) = n_{\theta, e} \, c_N^{-\frac34}\left( V_{\rm PQC}(\phi)  \over V_{\rm PQC}(\rho_e)  \right)^{\frac34}
\end{equation}
where $V_{\rm PQC}(\phi)$ is the inflaton energy density at the scale factor $a$ corresponding to  $\rho=\phi$.
Assuming that reheating is complete for $V_{\rm PQC}(\phi_{\rm RH})=g_* \pi^2 T_{RH}^4/30$, we obtain the PQ number abundance $Y_\theta$ at the reheating temperature $T_{RH}$ by
\bea
Y_{\theta} \equiv { n_{\theta}(T_{RH}) \over s(T_{RH})} \simeq 2.5\times 10^7 \, c_N^{-\frac34}\, \frac{n_{\theta,e}}{M_P^3} \,.\label{Ythend}
\eea

In our scenario for the KSVZ axion, the reheating process can occur due to the inflaton decays into the extra heavy quarks or RHNs,  $\phi \to QQ^c$ or $NN$,  through the Yukawa couplings in eq.~(\ref{KSVZ yukawa}), as far as they are allowed kinematically.
We assume that reheating is dominated by the decay process  $\phi \rightarrow QQ^c$, whereas the mass of RHN is already constrained by the cogenesis condition as in eq.~(\ref{yN}), setting a lower bound of the reheating temperature.  We note that the inflaton effective mass is $m_\phi(t) \equiv \sqrt{V''(\phi_0(t))}=\sqrt{3\lambda_\Phi} \phi_0(t)$ at a given scale (time) $a(t)$ for the quartic inflaton potential, whereas the masses of the fermions are  $m_f(t)=y_f\phi(t)/\sqrt{2}$ for $f=Q$ and $N$. Here, we note that $\phi(t) =\phi_0(t){\cal P}(t)$ where ${\cal P}(t)$ is a periodic function that crosses zero \cite{hmlee}. While the envelope of the inflaton field, $\phi_0(t)$ (``a slowly varing envelope''), red-shifts as $\phi_0(t) \propto 1/a(t)$, the highly oscillating modes of the inflaton with the frequency $E_n(t) \simeq 0.5 n m_\phi(t)$ can decay into fermions $f$ if $E_n(t) > 2 m_f(t)$. Furthermore, even the zero mode of the inflaton field can decay into fermions as the inflaton field approaches zero. Then, the time averaged decay rate of the inflaton over one oscillation  is expressed in terms of an ``effective coupling'' $y_{f,\rm eff}$ \cite{inflatonosc}:
\begin{equation}
    \Gamma_\phi(t) =\frac{N_f}{8\pi} \left({y_{f,\text{eff}} \over \sqrt{2}}\right)^2 m_\phi(t),\quad  
    y^2_{f,\text{eff}}\simeq 0.5 \, \alpha(\mathcal{R}) y_f^2,\quad 
    \mathcal{R} \equiv \left(\frac{4m_f}{m_\phi}\right)^2
\end{equation}
where $N_f$ is the number of degrees of freedom. 
In the two limits of $\mathcal{R}$, we have
\begin{equation}
    \alpha(\mathcal{R})\simeq 
    \begin{cases} 1  & \mathcal{R}\ll 1 \\
                0.5 \mathcal{R}^{-\frac12} \quad&  \mathcal{R}\gg 1
      \end{cases}
                \label{approx}
\end{equation}
For our analysis, we extend the first expression to ${\cal R}=1$, and the
second expression is reliable for ${\cal R}\geq 3$. In the intermediate region $1< {\cal R}<3$, we took $\alpha({\cal R})={\rm Min}(1,0.5{\cal R}^{-\frac{1}{2}})$.
Combining (\ref{approx}) with our approximate relation $\rho_{\rm RH}(T_{\rm RH}) = \rho_\phi(\phi_{\rm RH})\simeq{\lambda_\Phi\over4} \phi_{\rm RH}^4$, we get 
\begin{equation}
    T_{\rm RH} \simeq 1.5 \times 10^6 \mbox{ GeV} \left( {y_{f,\text{eff}} \over 10^{-4}} \right)^2  \left( {\lambda_\Phi \over 10^{-11}} \right)^{1/4},
\end{equation}
which agrees well with a rigorous analysis \cite{inflatonosc}.

\section{Parameter space for the PQ cogenesis}

In Fig. 1, we show regions of the successful cogenesis in the parameter space of the Yukawa coupling of the extra heavy quark $y_Q$ and the cutoff scale $\Lambda$ for two different values of $f_a=10^{8.6}$ GeV and $10^{8.9}$ GeV.
The orange and yellow regions represent ${\cal R}<1$ and ${\cal R}>1$, respectively. The maximum value of $y_Q$ is taken to be $10^{-3}$ above which the one-loop correction to $\lambda_\Phi$ becomes bigger than the tree-level value.  The region to the right of the left-upper  boundary of the yellow region is excluded because reheating cannot occur as the inflaton  settles down to the true vacuum, $\phi \simeq f_a$, and the decay channel is closed because $y_Q>\sqrt{\lambda_\Phi}$. The maximum reheating temperature and the corresponding field value $\phi_{\rm RH}$  are given by
\bea
T_{\rm RH}\approx 1.2\times 10^6 \left(\frac{M_P}{\Lambda}\right)^3 \text{ GeV},
\quad 
\phi_{\text{RH}} \approx 6.7\times 10^8 \left(\frac{M_P}{\Lambda}\right)^2 \text{ GeV},
\eea
respectively.
We note that $\phi_{\text{RH}}\ll \rho_c= \frac{M_P}{\sqrt{6}\zeta}$ in the region of our interest $\Lambda >10^{17}\,{\rm GeV}$.

\begin{figure}
    \centering
    \includegraphics[width=0.7\textwidth]{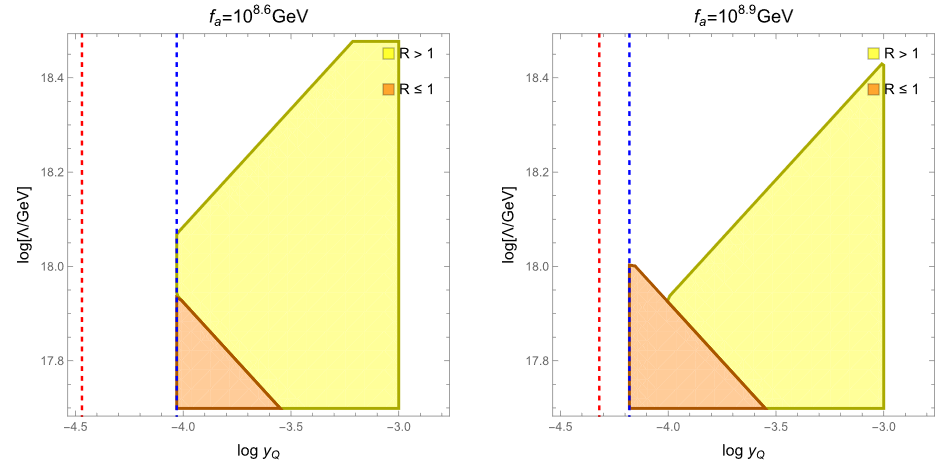}
    \caption{\raggedright Region of successful cogenesis in the parameter space ($y_Q$, $\Lambda$).  Reheating occurs in the region to the right of the red dashed line. The blue-dashed line shows  $y_Q=\sqrt{3}y_N$.  The boundary of the orange and yellow shaded regions corresponds to $\mathcal{R}=1$ }
    \label{Fig1}

\end{figure}
\begin{figure}
    \centering
    \includegraphics[width=0.75\textwidth]{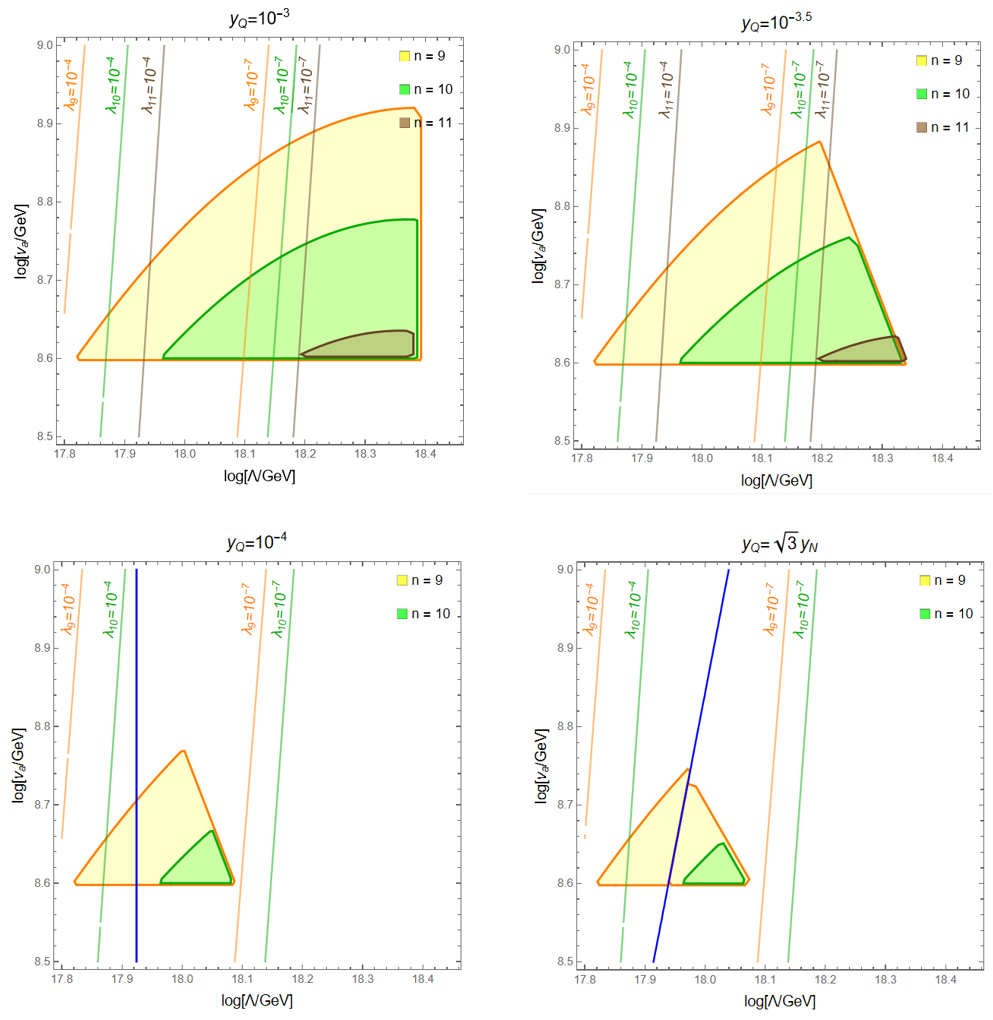}
    \caption{
    Parameter space for the successful cogenesis for different values of $y_Q$ in terms of  the cutoff scale $\Lambda$ vs. the PQ breaking scale $f_a$.  Allowed regions are indicated by orange, green, and brown, denoting PQ-violating potentials with dimensionalities $n=9$, 10,  and 11, respectively.  The required values for the coupling $\lambda_n$ are also depicted by almost vertical lines. The left (right) side of the blue line is for ${\cal R}<1$ ($>1$). 
    }
    \label{Fig2}
\end{figure}

Our final result is presented in Fig.~2 showing the regions of cogenesis in the parameter space of  $(\Lambda, f_a)$.  The allowed regions are found only for  PQ-violating potentials  $V_{\rm PQV}$ with dimensionalities $n=9,10$, and 11 corresponding to the areas in orange, green, and  brown, respectively.
The upper-left boundaries are determined by the inflation condition, $\frac{dV_{\rm PQC}}{d\phi}> \frac{dV_{\rm PQV}}{d\phi}$.    Inflaton decay to heavy quarks governs thee reheating process, which depends on their masses,  while the Yukawa coupling of the heavy quark is restricted to $\sqrt{3}y_N\leq y_Q \leq 10^{-3}$.  The blue line indicates ${\cal R}=1$; the region to its left (right) corresponds to ${\cal R}<1$ ($>1$). 
The upper-right boundaries are set by the condition whether the reheating can occur.  
We also find that the cogenesis  requires the values of the PQV couplings to be in the ranges of  
 $10^{-6.6} < \lambda_9 < 10^{-4.5}$ and $10^{-6.1} < \lambda_{10} < 10^{-5}$, for all the choices of $y_Q$ in Fig.~2, and 
  $10^{-9} \lesssim \lambda_{11} \lesssim 10^{-7}$ for $y_Q=10^{-3}, 10^{-3.5}$ in the upper panel of Fig.~2.

\section{Conclusion}

In this work, we have investigated a minimal extension of the KSVZ axion model to simultaneously explain the dark matter abundance and the baryon asymmetry of the Universe. To resolve the tension between the standard spontaneous baryogenesis mechanism and the overproduction of axion dark matter, we incorporated the seesaw mechanism into the Peccei-Quinn sector. This setup allows for spontaneous leptogenesis, where a $B-L$ asymmetry is generated near the seesaw scale and subsequently converted into the observed baryon asymmetry via electroweak sphalerons. We also addressed the origin of the initial kinetic misalignment required for this scenario by identifying the PQ field as the driver of inflation. Within the framework of pole inflation, the radial component of the PQ field plays the role of the inflaton, while the angular degree of freedom (the axion) acquires a large initial velocity through higher-dimensional PQ-breaking operators at the end of inflation. Our analysis demonstrates that the cogenesis of baryon asymmetry and dark matter is indeed viable in this framework.  Although the allowed region is found to be quite limited: $f_a=(4-9) \times 10^8$ GeV, our results suggest that the QCD axion, coupled with the seesaw mechanism and pole inflation, remains a compelling candidate for solving the strong CP problem while providing a unified origin for the matter content of the Universe.

\section*{Acknowledgments}

HML and JHS are supported in part by Basic Science Research Program through the National
Research Foundation of Korea (NRF) funded by the Ministry of Education, Science and
Technology (NRF-2022R1A2C2003567).

\section*{Appendix}
In this appendix, we consider the evolution of $\rho_\phi$ right after inflation.
To that, we solve the equation of motion of the non-canonical $\rho$ field numerically. From the Lagrangian in (\ref{Lagrangian E}), we get
\begin{figure}
    \centering
    \includegraphics[width=1\textwidth]{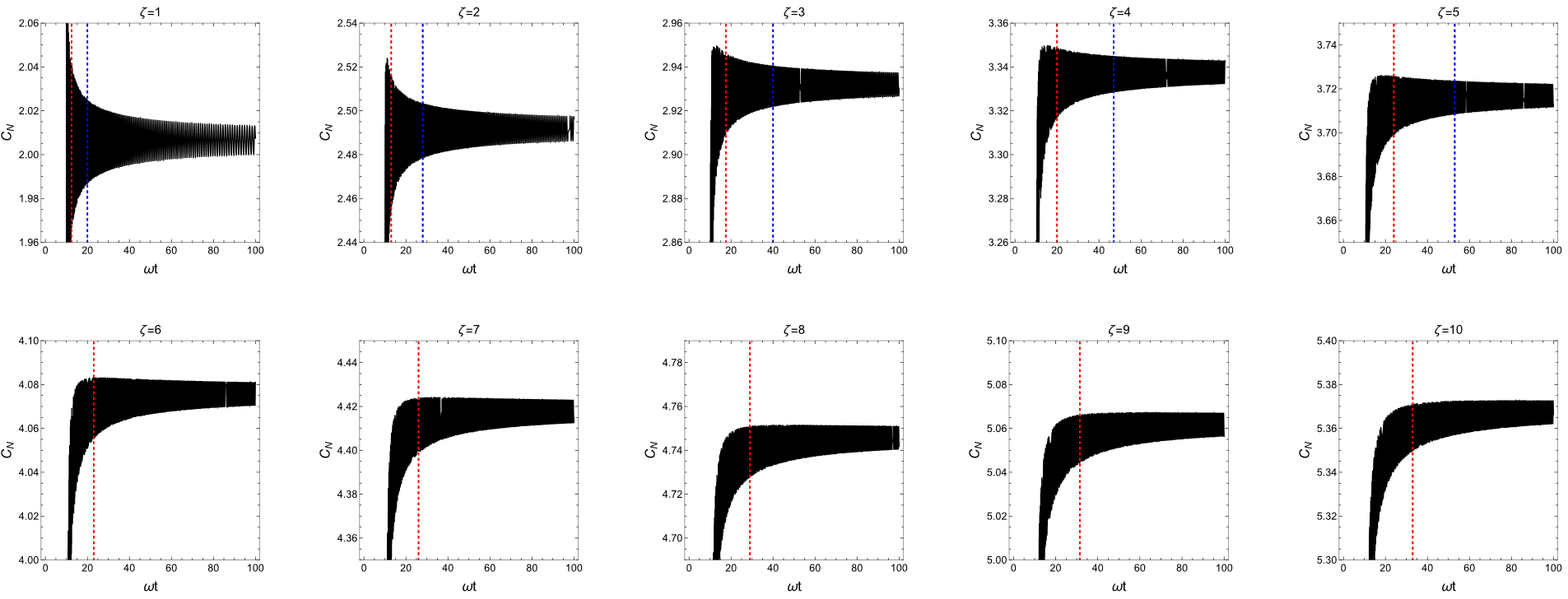}
    \caption{Time evolution of $c_N$ for $\zeta=1,2,\dots,10$, appearing in eq.~(\ref{Ythend}). The plots show that $c_N$ rapidly converges to a constant value for a given $\zeta$. Here, $wt=10$ with  $w=10^{11}$ GeV, and $t_e$ is the time at the end of inflation. The red and blues lines indicate the times when $\rho=\rho_c/10$ and  $\rho=\rho_c/20$, respectively.}
    \label{Fig3}
\end{figure}
\begin{align}
    \Ddot{\rho}=&-\left(\frac12 \frac{d\Xi}{d\rho}\frac{\Dot{\rho}^2}{\Xi(\rho)}+\frac{1}{\Xi(\rho)}\frac{dV_{\rm PQC}}{d\rho}+3H\Dot{\rho}\right) \label{e.o.m of rho}
\end{align}
where 
\begin{align}
\Xi(\rho)\equiv& \frac{1+\zeta(6\zeta-1)\frac{\rho^2}{M_P^2}}{\left(1-\zeta \frac{\rho^2}{M_P^2}\right)^2}.
\end{align}
At the end of inflation, the Hubble parameter is defined by
\begin{align}
    H^2=&\frac{1}{3M_P^2}\left[ \frac{1}{2}\Xi(\rho) \Dot{\rho}^2+V_{\rm PQC}(\rho) \right] \label{hubble constant}
\end{align}
where $H^2\approx V_{\rm PQC}/3M_P^2$.
Together  $\rho_e \approx 0.732/\sqrt{\zeta} $ and (28), we determine $\Dot{\rho}_e$ by using
\begin{align}
    \Dot{\rho}\approx -\frac{\partial V_{\rm PQC}}{\partial \rho}\left(\frac{\partial \rho}{\partial \phi}\right)^2 \frac{1}{3H},
\end{align}
which sets the initial condition for the numerical evolution in (\ref{e.o.m of rho}).
In Fig.~\ref{Fig3}, we show the asymptotic values of $c_N$ defined in (\ref{cN}) for $\zeta=1,2,\dots,10$, when $\rho \ll \rho_c=\frac{M_P}{\sqrt{6}\zeta}$. These reference times when $\rho=\rho_c/10$, $\rho_c/20$, indicate the onset of the regime where the non-canonical factor becomes $\Xi(\rho) \simeq 1$ and the background evolution is radiation-like. We find that the asymptotic value of $c_N$
 is well approximated by a linear function of $\zeta$, 
 \begin{align}
     c_N \simeq 0.37 \zeta+1.79,
 \end{align}
from which we get the numerical results for the PQ number within $\mathcal{O}(10 \%)$.


\begin{thebibliography}{99}

\bibitem{pq77}
R. D. Peccei and H. Quinn, Phys. Rev. Lett. 38, 1440 (1977); 
 Phys. Rev. D 16, 1791 (1977).

\bibitem{ww78}
 S. Weinberg, Phys. Rev. Lett. 40, 223 (1978); 
 F. Wilczek, ibid, 40, 279 (1978).

\bibitem{ksvz79}
 J.E. Kim, Phys. Rev. Lett. 43 (1979) 103;
 M.A. Shifman, A.I. Vainshtein and V.I. Zakharov,  Nucl. Phys. B 166 (1980) 493.

\bibitem{dfsz81}
M. Dine, W. Fischler and M. Srednicki, Phys. Lett. B 104 (1981) 199;
A.R. Zhitnitsky, Sov. J. Nucl. Phys. 31 (1980) 260.

\bibitem{cr24}
A.~Caputo and G.~Raffelt,
``Astrophysical Axion Bounds: The 2024 Edition,''
[arXiv:2401.13728 [hep-ph]].

\bibitem{misal83}
J. Preskill, M. B. Wise, and F. Wilczek, Phys. Lett. 120B, 127 (1983);
L. F. Abbott and P. Sikivie, Phys. Lett. 120B, 133 (1983);
M. Dine and W. Fischler, Phys. Lett. 120B, 137 (1983).

\bibitem{chh19}
R.~T.~Co, L.~J.~Hall and K.~Harigaya,
Phys. Rev. Lett. \textbf{124} (2020) no.25, 251802
[arXiv:1910.14152 [hep-ph]].

\bibitem{ck87}
A.~G.~Cohen and D.~B.~Kaplan,
Phys. Lett. B \textbf{199} (1987), 251-258; 
Nucl. Phys. B \textbf{308} (1988), 913-928.

\bibitem{ch19}
R.~T.~Co and K.~Harigaya,
Phys. Rev. Lett. \textbf{124} (2020) no.11, 111602
[arXiv:1910.02080 [hep-ph]].

\bibitem{seesaw77}
P. Minkowski, Phys. Lett. B 67 (1977), 421-428;
T. Yanagida, Conf. Proc. C 7902131 (1979), 95-99;
M. Gell-Mann, P. Ramond and R. Slansky, Conf. Proc. C 790927 (1979), 315-321;
[S. L. Glashow, NATO Sci. Ser. B 61 (1980), 687;
R. N. Mohapatra and G. Senjanovic, Phys. Rev. Lett. 44 (1980), 912.

\bibitem{cj23}
E.~J.~Chun and T.~H.~Jung,
Phys. Rev. D \textbf{109} (2024) no.9, 095004
[arXiv:2311.09005 [hep-ph]].

\bibitem{cls25}
E.~J.~Chun, H.~M.~Lee and J.~H.~Song,
[arXiv:2512.06413 [hep-ph]]

\bibitem{ad85}
I. Affleck and M. Dine
Nucl. Phy. B249 (1985) 361–380


\bibitem{hmlee}
H.~M.~Lee, A.~G.~Menkara, M.~J.~Seong and J.~H.~Song,
Eur. Phys. J. C \textbf{84} (2024) no.12, 1260
[arXiv:2408.17013 [hep-ph]];
H.~M.~Lee, A.~G.~Menkara, M.~J.~Seong and J.~H.~Song,
JHEP \textbf{05} (2024), 295
[arXiv:2310.17710 [hep-ph]].

\bibitem{bor16}
S.~Borsanyi, Z.~Fodor, J.~Guenther, K.~H.~Kampert, S.~D.~Katz, T.~Kawanai, T.~G.~Kovacs, S.~W.~Mages, A.~Pasztor and F.~Pittler, \textit{et al.}
Nature \textbf{539} (2016) no.7627, 69-71
[arXiv:1606.07494 [hep-lat]].


\bibitem{DEMY20}
V.~Domcke, Y.~Ema, K.~Mukaida and M.~Yamada,
``Spontaneous Baryogenesis from Axions with Generic Couplings,''
JHEP \textbf{08} (2020), 096
[arXiv:2006.03148 [hep-ph]].



\bibitem{pdg}
S.~Navas \textit{et al.} [Particle Data Group],
Phys. Rev. D \textbf{110} (2024) no.3, 030001
doi:10.1103/PhysRevD.110.030001


\bibitem{pole}
R.~Kallosh and A.~Linde,
JCAP \textbf{12} (2013), 006
[arXiv:1309.2015 [hep-th]];
R.~Kallosh, A.~Linde and D.~Roest,
JHEP \textbf{11} (2013), 198
[arXiv:1311.0472 [hep-th]];
M.~Galante, R.~Kallosh, A.~Linde and D.~Roest,
Phys. Rev. Lett. \textbf{114} (2015) no.14, 141302
[arXiv:1412.3797 [hep-th]];
B.~J.~Broy, M.~Galante, D.~Roest and A.~Westphal,
JHEP \textbf{12} (2015), 149
[arXiv:1507.02277 [hep-th]];
T.~Terada,
Phys. Lett. B \textbf{760} (2016), 674-680
[arXiv:1602.07867 [hep-th]].


\bibitem{pole2}
S.~Clery, H.~M.~Lee and A.~G.~Menkara,
JHEP \textbf{10} (2023), 144
[arXiv:2306.07767 [hep-ph]].


\bibitem{axionquality}
M.~Kamionkowski and J.~March-Russell,
Phys. Lett. B \textbf{282} (1992), 137-141
[arXiv:hep-th/9202003 [hep-th]];
S.~M.~Barr and D.~Seckel,
Phys. Rev. D \textbf{46} (1992), 539-549

\bibitem{inflatonosc}
M.~A.~G.~Garcia, K.~Kaneta, Y.~Mambrini and K.~A.~Olive,
JCAP \textbf{04} (2021), 012
[arXiv:2012.10756 [hep-ph]].

\end{thebibliography}
\end{document}